\documentclass[12pt]{iopart}

\usepackage{epsfig}

\def\d{{\rm d}}
\def\p{I\!\!P}

\def\lr{\left( }
\def\rr{\right) }
\def\le{\left[ }
\def\re{\right] }

\def\beq{\begin{equation}}
\def\eeq{\end{equation}}
\def\bea{\begin{eqnarray}}
\def\eea{\end{eqnarray}}

\begin{document}

\vspace*{-2cm}
\noindent DESY 05-095\hfill ISSN 0418-9833 \\
LPSC 05-053 \\
hep-ph/0506121 \\
\vspace*{1cm}

\title[Factorization scheme and scale dependence in diffractive dijet
 production ...]{Factorization scheme and scale dependence in diffractive
 dijet production at low $Q^2$}

\author{Michael Klasen\dag\ \footnote[3]{klasen@lpsc.in2p3.fr} and Gustav Kramer\ddag  
}

\address{\dag\ Laboratoire de Physique Subatomique et de Cosmologie,
 Universit\'e Joseph Fourier/CNRS-IN2P3, 53 Avenue des Martyrs, F-38026
 Grenoble, France}

\address{\ddag\ II.\ Institut f\"ur Theoretische Physik, Universit\"at
 Hamburg, Luruper Chaussee 149, D-22761 Hamburg, Germany}

\begin{abstract}
We calculate diffractive dijet production in deep-inelastic scattering at
next-to-leading order of perturbative QCD, including contributions
from direct and resolved photons, and compare our predictions to preliminary
data from the H1 collaboration at HERA. We study how the cross section
depends on the factorization scheme and scale $M_{\gamma}$ at the virtual
photon vertex for the occurrence of factorization breaking. The strong
$M_{\gamma}$-dependence, which is present when only the resolved cross
section is suppressed, is tamed by introducing the suppression also into the
initial-state NLO correction of the direct part.
\end{abstract}



\maketitle

\section{Introduction}
\label{sec:1}

\begin{figure}
 \centering
 \epsfig{file=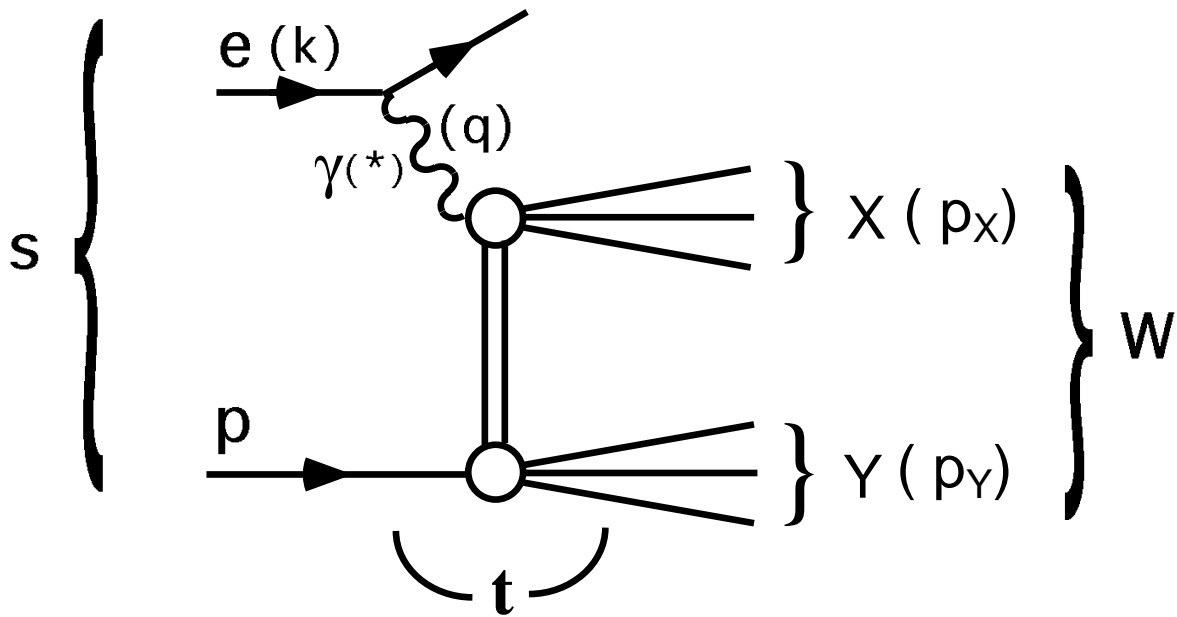,width=\columnwidth}
 \caption{\label{fig:1}Diffractive scattering process $ep\to eXY$, where
 the hadronic systems $X$ and $Y$ are separated by the largest rapidity
 gap in the final state.}
\end{figure}

It is well known that in high-energy deep-inelastic $ep$-collisions a large
fraction of the observed events are diffractive. These events are defined
experimentally by the presence of a forward-going system $Y$ with
four-momentum $p_Y$, low mass $M_Y$ (in most cases a single proton or a
proton plus low-lying nucleon resonances), small momentum transfer squared
$t=(p-p_Y)^2$, and small longitudinal momentum transfer fraction
$x_{\p}=q(p-p_Y)/qp$ from the incoming proton with four-momentum $p$ to the
system $X$ (see Fig.\ \ref{fig:1}). The presence of a hard scale, as for
example the photon virtuality $Q^2=-q^2$ in deep-inelastic scattering (DIS)
or the large transverse jet momentum $p_T^{*}$ in the photon-proton
center-of-momentum frame, should then allow for calculations of the
production cross section for the central system $X$ with the known methods
of perturbative QCD. Under this assumption, the cross section for the
inclusive production of two jets, $e+p \rightarrow e+2jets+X'+Y$, can be
calculated from the  well-known formul\ae\ for jet production in
non-diffractive $ep$ collisions, where in the convolution of the partonic
cross section with the parton distribution functions (PDFs) of the proton
the latter ones are replaced by the diffractive PDFs. In the simplest
approximation, they are described by the exchange of a single, factorizable
pomeron/Regge-pole. \\

The diffractive PDFs have been determined by the H1 collaboration at HERA
from high-precision inclusive measurements of the DIS process $ep\rightarrow
eXY$ using the usual DGLAP evolution equations in leading order (LO) and
next-to-leading order (NLO) and the well-known formula for the inclusive
cross section as a convolution of the inclusive parton-level cross section
with the diffractive PDFs \cite{h1ichep02}. For a similar analysis of the
inclusive measurements of the ZEUS collaboration see \cite{Chekanov:2004hy}.
For inclusive diffractive DIS it has been proven by Collins that the
formula referred to above is applicable without additional corrections and
that the inclusive jet production cross section for large $Q^2$ can be
calculated in terms of the same diffractive PDFs \cite{Collins:1997sr}. The
proof of this factorization formula, usually referred to as the validity of
QCD factorization in hard diffraction, also appears to be valid for the
direct part of photoproduction ($Q^2\simeq0$) or low-$Q^2$ electroproduction
of jets \cite{Collins:1997sr}. However, factorization does not hold for hard
processes in diffractive hadron-hadron scattering. The problem is that soft
interactions between the ingoing two hadrons and their remnants occur in
both the initial and final state. This agrees with experimental measurements
at the Tevatron \cite{Affolder:2000vb}. Predictions of diffractive dijet
cross sections for $p\bar{p}$ collisions as measured by CDF using the same
PDFs as determined by H1 \cite{h1ichep02} overestimate the measured cross
section by up to an order of magnitude \cite{Affolder:2000vb}. This
suppression of the CDF cross section can be explained by considering the
rescattering of the two incoming hadron beams which, by creating additional
hadrons, destroy the rapidity gap \cite{Kaidalov:2001iz}.\\

%
%

Processes with real photons ($Q^2 \simeq 0$) or virtual photons with fixed,
but low $Q^2$ involve direct interactions of the photon with quarks from the
proton as well as resolved photon contributions, leading to parton-parton
interactions and an additional remnant jet coming from the photon (for a
review see \cite{Klasen:2002xb}). As already said, factorization should be
valid for direct interactions as in the case of DIS, whereas it is expected
to fail for the resolved process similar as in the hadron-hadron scattering
process. In a two-channel eikonal model similar to the one used to calculate
the suppression factor in hadron-hadron processes \cite{Kaidalov:2001iz},
introducing vector-meson dominated photon fluctuations, a suppression by
about a factor of three for resolved photoproduction at HERA is predicted
\cite{Kaidalov:2003xf}. Such a suppression factor has recently been applied
to diffractive dijet photoproduction \cite{Klasen:2004tz,Klasen:2004qr} and
compared to preliminary data from H1 \cite{h1ichep04} and ZEUS
\cite{zeusichep04}. While at LO no suppression of the resolved contribution
seemed to be necessary, the NLO corrections increase the cross section
significantly, showing that factorization breaking occurs at this order for
resolved photoproduction and that a suppression factor $R=0.34$, in
agreement with the prediction of \cite{Kaidalov:2003xf}, gives a reasonable
description of the experimental data of \cite{h1ichep04} and
\cite{zeusichep04}. \\

As already mentioned in our earlier work \cite{Klasen:2004tz,Klasen:2004qr},
describing the factorization breaking in hard photoproduction as well as in
electroproduction at very low $Q^2$ \cite{KK} by suppressing the resolved
contribution only may be problematic. An indication for this is the fact
that the separation between the direct and the resolved process is uniquely
defined only in LO. In NLO these two processes are related. The separation
depends on the factorization scheme and the factorization scale
$M_{\gamma}$. The sum of both cross sections is the only physically relevant
cross section, which is approximately independent of the factorization
scheme and scale \cite{BKS}. We demonstrated already in
\cite{Klasen:2004tz,Klasen:2004qr} that by multiplying the resolved cross
section with the suppression factor $R=0.34$, the correlation of the
$M_{\gamma}$-dependence between the direct and resolved part is destroyed
and the sum of both parts has a stronger $M_{\gamma}$-dependence than for
the unsuppressed case ($R=1$), where the $M_{\gamma}$-dependence of the NLO
direct cross section is compensated to a high degree against the
$M_{\gamma}$-dependence of the LO resolved part. The introduction of the
resolved cross section is dictated by perturbation theory. At NLO, collinear
singularities arise from the photon initial state, which are absorbed at the
factorization scale into the photon PDFs. This way the photon PDFs become
$M_{\gamma}$-dependent. The equivalent $M_{\gamma}$-dependence, just with
the opposite sign, is left in the NLO corrections to the direct
contribution. With this knowledge, it is obvious that we can obtain a
physical cross section at NLO, {\it i.e.} the superposition of the NLO
direct and LO resolved cross section, with a suppression factor $R<1$ and no
$M_{\gamma}$-dependence left, if we also multiply the
$\ln M_{\gamma}$-dependent term of the NLO correction to the direct
contribution with the same suppression factor as the resolved cross
section.\\

It is the purpose of this paper to present the special form of the
$\ln M_{\gamma}$-term in the NLO direct contribution and then, after
multiplying it with the suppression factor $R$, to demonstrate that the
$M_{\gamma}$-dependence of the physical cross section cancels to a large
extent in the same way as in the unsuppressed case ($R=1$), if we modify the
suppression mechanism in the way as explained above. Of course, some small
$M_{\gamma}$-dependence will remain, also in the case with the suppression
of the $\ln M_{\gamma}$-part in the direct contribution, which is due to
the evolution of the photon PDFs and/or the NLO correction to the resolved
part, which is left uncompensated as long as the NNLO contribution to the
direct part is not included.\\

In addition to checking the $M_{\gamma}$ scale dependence we shall study how
the NLO cross section depends on the factorization scheme for the photon
PDFs. These studies can be done for photoproduction ($Q^2 \simeq 0$) as well
as for electroproduction with fixed, small $Q^2$. Since in electroproduction
the initial-state singularity in the limit $Q^2 \rightarrow 0$ is more
directly apparent than for the photoproduction case, where this singularity
appears as a pole in $d-4$ using dimensional regularization, we shall
consider in this paper the low-$Q^2$ electroproduction case just for
demonstration. This diffractive dijet cross section has been calculated
recently \cite{KK}. In this work $\d\sigma/\d Q^2$ as a function of $Q^2$
and $\d\sigma/\d z_{\p}$ as a function of $z_{\p}$ for various
$Q^2+p^{*2}_T$ bins was calculated for H1 kinematical conditions and
compared to the preliminary experimental data from H1
\cite{Schatzel:2004be}. Here, $z_{\p}$ is the parton momentum fraction in
the pomeron. In the following we shall consider these cross sections with
the same kinematical constraints as in \cite{KK}.\\

A consistent factorization scheme for low-$Q^2$ virtual photoproduction
has been defined and the full (direct and resolved) NLO corrections for
inclusive dijet production have been calculated in \cite{Klasen:1997jm}. In
this work we adapt this inclusive NLO calculational framework to diffractive
dijet production at low-$Q^2$ in the same way as in \cite{KK}, except that
we multiply the $\ln M_{\gamma}$ dependent terms as well as the resolved
contributions with the same suppression factor $R=0.34$ as in our earlier
work \cite{Klasen:2004tz,Klasen:2004qr,KK}. The exact value of this
suppression factor may change in the future, when better data for
photoproduction and low-$Q^2$ electroproduction have been analyzed.\\

In the next Section, we shall define the H1 kinematical constraints and
present the $\ln M_{\gamma} $-dependent terms in the NLO corrections to
the direct cross section, as they are reported in \cite{Klasen:1997jm}. We
shall also define which specific contribution of this initial-state NLO
correction will be suppressed with the same factor $R$ as the resolved part.
Then we give our results concerning the scheme dependence of the photon PDFs
by looking at $\d\sigma/\d Q^2$ and $\d\sigma/\d z_{\p}$ using two different
schemes. After this we present the $\ln M_{\gamma}$-dependence of the partly
suppressed NLO direct and the fully suppressed NLO resolved cross section
$\d\sigma/\d Q^2$ and their sum for the lowest $Q^2$ bin. After this we
study the $M_{\gamma}$-dependence of $\d\sigma/\d Q^2$ for the other
$Q^2$-bins and of $\d\sigma/\d z_{\p}$ and compare always with the H1 data.
As the last point we study the contribution of the hadron-like or
vector-dominance part of the resolved cross section in order to see whether
the suppression of this part alone could be an alternative to solve the
$M_{\gamma}$ scale problem as suggested in
\cite{Klasen:2004tz,Klasen:2004qr}.

\section{Kinematical constraints and results}
\label{sec:2}

The factorization scheme for virtual photoproduction has been defined and
the full NLO corrections for inclusive dijet production have been calculated
in \cite{Klasen:1997jm}. They have been implemented in the NLO Monte Carlo
program JET\-VIP \cite{Potter:1999gg}. We have adapted this NLO framework to
diffractive dijet production. According to \cite{Klasen:1997jm}, the
subtraction term, which is absorbed into the PDFs of the virtual photon
$f_{a/\gamma}(x,M_{\gamma})$, is of the form
\bea
 S_{q_i \leftarrow \gamma}(z,M_{\gamma}) &=& \ln\lr\frac{M_{\gamma}^2}
 {Q^2(1-z)}\rr P_{q_i \leftarrow \gamma}(z) -N_c Q_i^2. \label{eq:1}
\eea
In Eq. (\ref{eq:1}), $z=p_1p_2/p_0q \in [x;1]$ and the splitting function
$P_{q_i \leftarrow \gamma} (z)$ is
\beq
 P_{q_i \leftarrow \gamma}(z) = 2 N_c Q_i^2 \frac{z^2+(1-z)^2}{2}.
\eeq
$Q_i$ is the fractional charge of the quark $q_i$, $p_1$ and $p_2$ are the
momenta of the two outgoing jets, and $p_0$ and $q$ are the momenta of the
ingoing parton and virtual photon, respectively. Since $Q^2\ll
M_{\gamma}^2$, the subtraction term in Eq.\ (\ref{eq:1}) is large and is
therefore resummed by the DGLAP evolution equations for the virtual photon
PDFs. After the subtraction of Eq.\ (\ref{eq:1}), the finite term, which
remains in the matrix element for the NLO correction to the direct process,
is \cite{Klasen:1997jm}
\bea
 M(Q^2)_{\overline{\rm MS}} &=& -\frac{1}{2N_c} P_{q_i \leftarrow \gamma}(z)
 \ln \lr\frac{M_{\gamma}^2 z}{(zQ^2+y_ss)(1-z)}\rr
 +{Q_i^2\over2}. \label{eq:3}
\eea
This term is defined by the requirement that it is equal to
$M_{\overline{\rm MS}}$ for real photons, if $Q^2=0$. For our study of the
$M_{\gamma}$-dependence of the physical cross section it is essential to
recognize that the $M_{\gamma}$-dependence in Eq. (\ref{eq:3}) is the same
as in the subtraction term Eq. (\ref{eq:1}), {\it i.e.} $\ln M_{\gamma}$ is
multiplied with the same factor (except that the subtraction term $S$ has to
be multiplied by $-1/(2N_c)$). As already mentioned, this yields the
$M_{\gamma}$-dependence before the evolution is turned on. In the usual
non-diffractive dijet photoproduction these two $M_{\gamma}$-dependences,
{\it i.e.} in Eqs.\ (\ref{eq:1}) and (\ref{eq:3}), cancel, when the NLO
correction to the direct part is added to the LO resolved cross section
\cite{BKS}. Then it is obvious that the approximate
$M_{\gamma}$-independence is destroyed, if the resolved cross section is
multiplied by the suppression factor $R$ to account for the factorization
breaking in the experimental data. To remedy this deficiency, we propose to
multiply the $\ln M_{\gamma}$-dependent term in $M(Q^2)_{\overline{\rm MS}}$
with the same suppression factor as the resolved cross section. This is done
in the following way: We split $M(Q^2)_{\overline{\rm MS}}$ into two terms
using the scale $p_T^{*}$ in such a way that the term containing the slicing
parameter $y_s$, which was used to separate the initial-state singular
contribution, remains unsuppressed. In particular, we replace
$M(Q^2)_{\overline{\rm MS}}$ by
\bea
 M(Q^2,R)_{\overline{\rm MS}} &=& \le-\frac{1}{2N_c} P_{q_i\leftarrow
 \gamma}(z)\ln\lr\frac{M_{\gamma}^2 z}{p_T^{*2}(1-z)}\rr+{Q_i^2\over2} \re R
 \nonumber \\
 && \ -\frac{1}{2N_c} P_{q_i\leftarrow\gamma}(z)
 \ln\lr\frac{p_T^{*2}}{zQ^2+y_s s}\rr,\label{eq:4}
\eea
where $R$ is the suppression factor. This expression coincides with
$M(Q^2)_{\overline{\rm MS}}$ in Eq.\ (\ref{eq:3}) for $R=1$, as it should,
and leaves the second term in Eq.\ (\ref{eq:4}) unsuppressed. In Eq.\
(\ref{eq:4}) we have suppressed in addition to $\ln(M_{\gamma}^2/p_T^{*2})$
also the $z$-dependent term $\ln (z/(1-z))$, which is specific to the
$\overline{\rm MS}$ subtraction scheme as defined in \cite{Klasen:1997jm}.
To keep this term unsuppressed, {\it i.e.} to move it to the second term in
Eq.\ (\ref{eq:4}), would again produce an inconsistency with the suppressed
resolved contribution. The second term in Eq.\ (\ref{eq:4}) must be left in
its original form, {\it i.e.} being unsuppressed, in order to achieve the
cancellation of the slicing parameter dependence of the complete NLO
correction in the limit of very small $Q^2$ or equivalently very large $s$.
With these arguments the splitting of the terms in Eq.\ (\ref{eq:3}), which
can be suppressed, and those terms, which remain unsuppressed, is almost
unique. It is clear that the suppression of this part of the NLO correction
to the direct cross section will change the full cross section only very
little as long as we choose $M_{\gamma} \simeq p_T^{*}$. The first term in
Eq.\ (\ref{eq:4}), which has the suppression factor $R$, will be denoted by
${\rm DIR}_{\rm IS}$ in the following. \\

To study the left-over $M_{\gamma}$-dependence of the physical cross
section, we have calculated the diffractive dijet cross section with the
following kinematical constraints as in the H1 experiment
\cite{Schatzel:2004be}: The electron and proton beam energies are 27.5 and
820 GeV, respectively, $4<Q^2<80$ GeV$^2$ is the range of the squared
electron momentum transfer, $0.1<y<0.7$ is the interval of virtual photon
energy fraction, $x_{\p}<0.05$, $|t|<1$ GeV$^2$, $M_Y<1.6$ GeV,
$p_{T,jet1,2}^{*} > 5(4)$ GeV, and jet rapidities $-3 <\eta_{jet1,2}^{*}<0$.
Jets are defined by the CDF cone algorithm with jet radius equal to one. The
asymmetric cuts for the transverse momenta of the two jets are required for
infrared stable comparisons with the NLO calculations \cite{Klasen:1995xe}.
The original H1 analysis actually used a symmetric cut of 4 GeV on the
transverse momenta of both jets \cite{Adloff:2000qi}. The data have,
however, been reanalyzed for asymmetric cuts \cite{Schatzel:2004be}. In the
extraction of the diffractive parton densities the usual Regge factorization
ansatz \cite{Ingelman:1984ns}
\bea
 f_i^D(\xi,Q^2,x_{\p},t)&=&f_{\p/p}(x_{\p},t) f_i\lr\frac{\xi}{x_{\p}},Q^2
 \rr\label{eq:5}
\eea
has been employed, which we shall use also in our computations. This ansatz
neglects a possible scale dependence of $f_{\p/p}$, which could be
responsible for a change of the $x_{\p}$ dependence between the soft and the
hard pomeron exchange. In the analysis of H1 \cite{h1ichep02} the intercept
$\alpha_{\p}(0)$ is extracted from the diffractive DIS data. For an
alternative approach, where the factorization assumption in Eq.\
(\ref{eq:5}) is modified see \cite{Martin:2004xw}.\\

For the NLO resolved virtual photon predictions, we have used the PDFs SaS1D
\cite{Schuler:1996fc} and transformed them from the DIS$_{\gamma}$ to the 
$\overline{\rm MS}$ scheme as in \cite{Klasen:1997jm}. If not stated
otherwise, the renormalization and factorization scales at the pomeron and
the photon vertex are equal and fixed to $p_T^{*} = p_{T,jet1}^{*}$. We
include four flavours, {\it i.e.} $n_f=4$ in the formula for $\alpha_s$ and
in the PDFs of the pomeron and the photon. The SaS-type PDFs for the virtual
photon are chosen, since in their parametrization the scale dependence
following from the evolution is given separately for the hadronic part and
the point-like part, which is called the anomalous part in
\cite{Schuler:1996fc}. \\

\begin{figure}
 \centering
 \epsfig{file=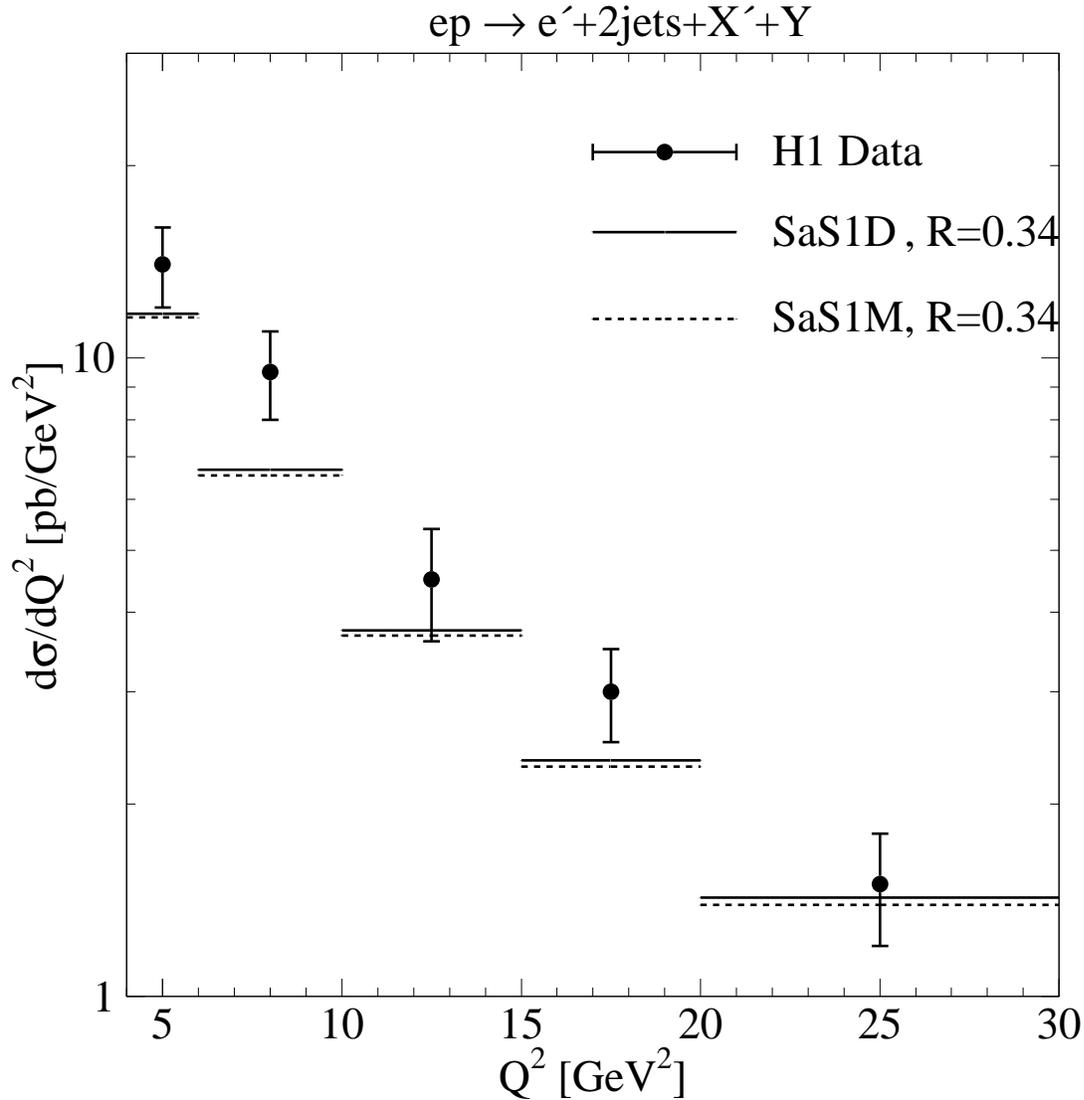,width=\columnwidth}
 \caption{\label{fig:2} Dependence of the diffractive dijet cross section at
 HERA on the squared photon virtuality $Q^2$. Preliminary H1 data
 \cite{Schatzel:2004be} are compared with theoretical NLO predictions
 including resolved virtual photon and direct contributions with a
 suppression factor of $R=0.34$ for the resolved and for the DIR$_{\rm IS}$
 term using virtual photon PDFs SaS1D (full) and SaS1M (dashed) from
 \cite{Schuler:1996fc}.}
\end{figure}

\begin{figure}
 \centering
 \epsfig{file=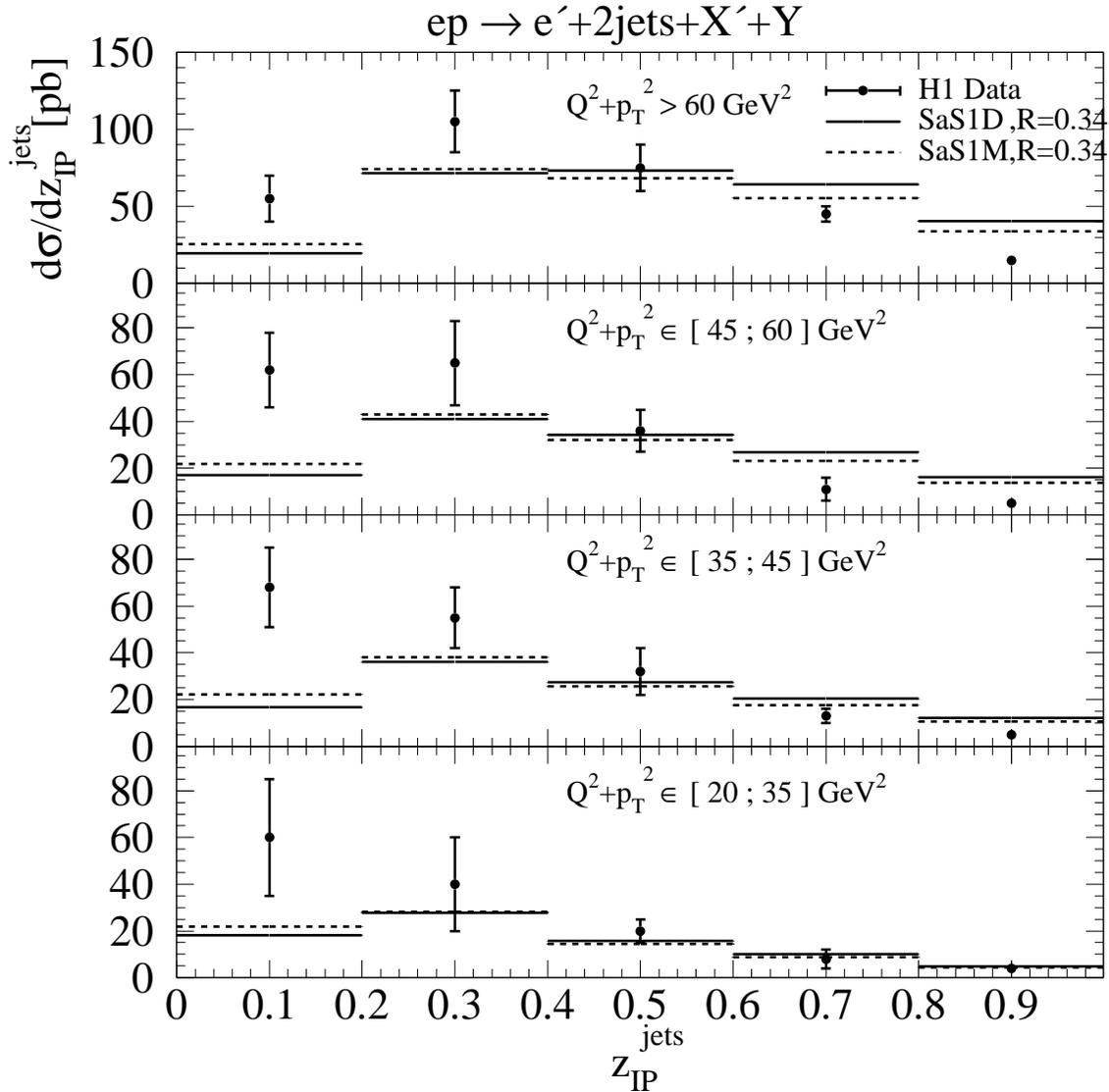,width=\columnwidth}
 \caption{\label{fig:3}Dependence of the diffractive dijet cross section at
 HERA on the parton momentum fraction in the pomeron $z_{\p}$ for different
 ranges of $Q^2+p_T^{*2}$. Preliminary H1 data \cite{Schatzel:2004be} are
 compared with two theoretical predictions (as in Fig.\ \ref{fig:2}) using
 SaS1D (full) and SaS1M (dashed) virtual photon PDFs \cite{Schuler:1996fc}.}
\end{figure}

\begin{figure}
 \centering
 \epsfig{file=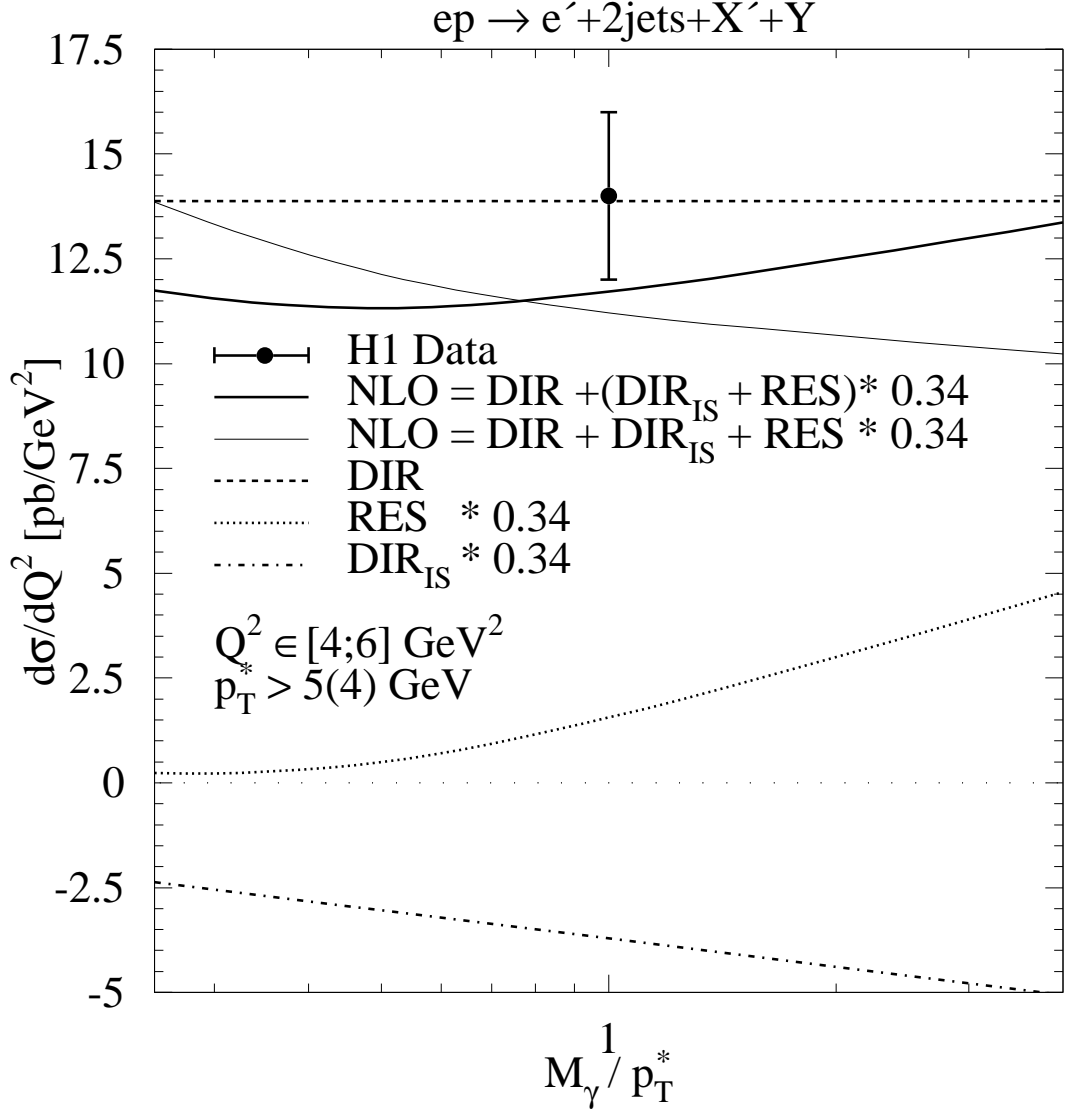,width=\columnwidth}
 \caption{\label{fig:4}Photon factorization scale dependence of resolved and
 direct contributions to $\d\sigma/\d Q^2$ together with their weighted sums
 for (i) suppression of the resolved cross section and for (ii) additional 
 suppression of DIR$_{\rm IS}$, using SaS1D virtual photon PDFs
 \cite{Schuler:1996fc}.}
\end{figure}

With these assumptions we have calculated the same cross section as in our
previous work \cite{KK}. First we investigated how the cross section
$\d\sigma/\d Q^2$ depends on the factorization scheme of the PDFs for the
virtual photon, {\it i.e.} $\d\sigma/\d Q^2$ is calculated for the choice
SaS1D and SaS1M. Here $\d\sigma/\d Q^2$ is the full cross section (sum of
direct and resolved) integrated over the momentum and rapidity ranges
referred to above. The results are shown in Fig.\ \ref{fig:2}, where
%
%
$\d\sigma/\d Q^2$ with SaS1D (full line) is compared to $\d\sigma/\d Q^2$
with SaS1M (dashed line). The theoretical cross sections are integrated over
the same $Q^2$ bins as the experimental ones. In this figure and in all the
following ones, except Fig.\ \ref{fig:4}, where we study the explicit
$M_{\gamma}$-dependence, both predictions are obtained with suppressed
resolved cross section and the additional suppression of DIR$_{\rm IS}$
as described above with suppression factor $R=0.34$. As we can see in Fig.\
\ref{fig:2}, the choice of the factorization scheme of the virtual photon
PDFs has negligible influence on $\d\sigma/\d Q^2$ for all considered $Q^2$.
The predictions agree reasonably well with the preliminary H1 data
\cite{Schatzel:2004be}. Comparing with our previous results \cite{KK} 
we notice that the cross sections in Fig.\ \ref{fig:2} are somewhat smaller
than the corresponding ones in Fig.\ 2 of \cite{KK}. This is, however,
{\it not} the effect of the additional suppression of a NLO term in the
direct cross section. The difference comes instead from a different choice
of the overall scales in \cite{KK} which was fixed to the average $\langle
p_T^{*}\rangle = \sqrt{40}$ GeV. Another check on the effect of a different
choice of the factorization scheme is shown in Fig.\ \ref{fig:3}, where the
dependence of
%
%
the cross section $\d\sigma/\d z_{\p}$ as a function of $z_{\p}$ is plotted
for various ranges of $Q^2+p_T^{*2}$. From this plot we see that the scheme
dependence is again very small for all $Q^2+p_T^{*2}$ ranges and for all
$z_{\p}\in [0;1]$. Except for the lowest $z_{\p}$ bin and for the two
highest $z_{\p}$ bins in the two upper $Q^2+p_T^{*2}$ ranges, we observe
reasonable agreement with the data. The agreement is best for the smallest
$Q^2+p_T^{*2}\in [20;35]~GeV^2$ region and $z_{\p} \geq 0.2$.\\

We now turn to the $M_{\gamma}$-dependence of the cross section with a
suppression factor for DIR$_{\rm IS}$, which is the main part of this paper.
To show this dependence for the two suppression mechanisms, (i) suppression
of the resolved cross section only and (ii) additional suppression of the 
DIR$_{\rm IS}$ term as defined in Eq.\ (\ref{eq:4}) in the NLO correction of
the direct cross section, we consider $\d\sigma/\d Q^2$ for the lowest
$Q^2$-bin, $Q^2\in [4,6]$ GeV$^2$. In Fig.\ \ref{fig:4}, this cross section
%
%
is plotted as a function of $\xi=M_{\gamma}/p_T^{*}$ in the range $\xi\in
[0.25,4]$ for the cases (i) (light full curve) and (ii) (full curve). We see
that the cross section for case (i) has an appreciable $\xi$-dependence in
the considered $\xi$ range of the order of $40\%$, which is caused by the
suppression of the resolved contribution only. With the additional
suppression of the DIR$_{\rm IS}$ term in the direct NLO correction, the
$\xi$-dependence of $\d\sigma/\d Q^2$ is reduced to approximately less than
$20\%$, if we compare the maximal and the minimal value of $\d\sigma/\d Q^2$
in the considered $\xi $ range. The remaining $\xi $-dependence is caused by
the NLO corrections to the suppressed resolved cross section and the
evolution of the virtual photon PDFs. How the compensation of the
$M_{\gamma}$-dependence between the suppressed resolved contribution and the
suppressed direct NLO term works in detail is exhibited by the dotted and
dashed-dotted curves in Fig.\ \ref{fig:4}. The suppressed resolved term
increases and the suppressed direct NLO term decreases by approximately the
same amount with increasing $\xi$. In addition we show also $\d\sigma/\d
Q^2$ in the DIS theory, {\it i.e.} without subtraction of any $\ln Q^2$
terms (dashed line). Of course, this cross section must be independent of
$\xi$. This prediction agrees very well with the experimental point, whereas
the result for the subtracted and suppressed theory (full curve) lies
slightly below. We notice, that for $M_{\gamma}=p^{*}_T$ the additional
suppression of DIR$_{\rm IS}$ has only a small effect. It increases
$\d\sigma/\d Q^2$ by $5\%$ only.\\

\begin{figure}
 \centering
 \epsfig{file=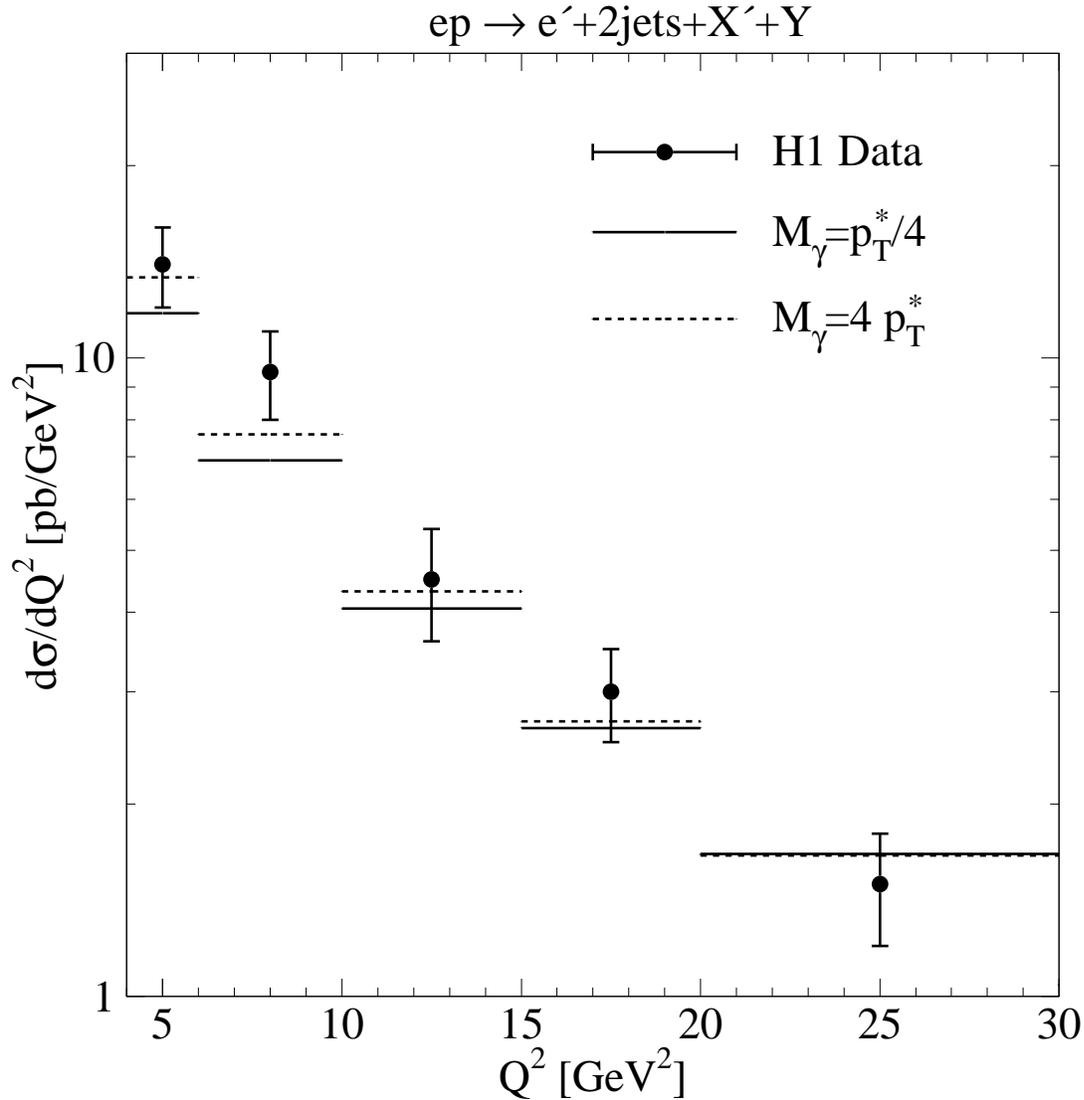,width=\columnwidth}
 \caption{\label{fig:5}Same as in Fig.\ \ref{fig:2} for $M_{\gamma}=
 p^{*}_T/4$ (full)and $M_{\gamma}=4p^{*}_T$ (dashed) and comparison with
 preliminary H1 data using SaS1D virtual photon PDFs \cite{Schuler:1996fc}.}
\end{figure}

\begin{figure}
 \centering
 \epsfig{file=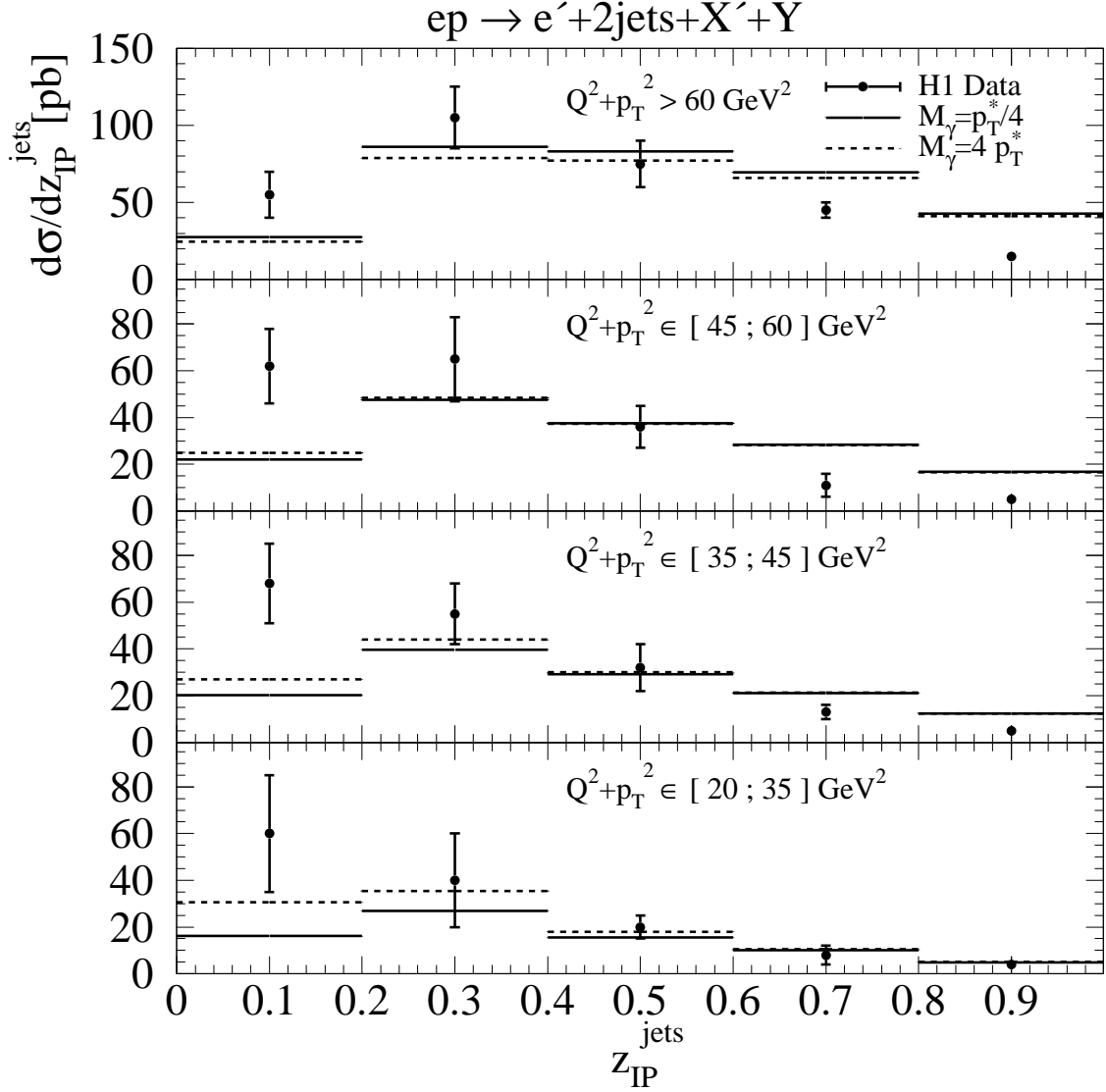,width=\columnwidth}
 \caption{\label{fig:6}Same as in Fig.\ \ref{fig:3} for $M_{\gamma}=
 p^{*}_T/4$ (full) and $M_{\gamma}=4p^{*}_T$ (dashed) and comparison with
 preliminary H1 data using SaS1D virtual photon PDFs \cite{Schuler:1996fc}.}
\end{figure}

In order to get an idea about the $M_{\gamma}$ scale dependence of 
$\d\sigma/\d Q^2$ for the other $Q^2$ bins and for $\d\sigma/\d z_{\p}$ in
the $Q^2+p_T^{*2}$ ranges as in Fig.\ \ref{fig:3}, we have computed these
two cross sections for two choices of $M_{\gamma}$, namely $M_{\gamma}=
p_T^{*}/4$ and $M_{\gamma}=4p_T^{*}$ corresponding to the lowest and highest
$\xi$ in Fig.\ \ref{fig:4}. The result for the $\d\sigma/\d Q^2$ is shown in
Fig.\ \ref{fig:5}. We see that the $M_{\gamma}$-dependence in the considered
%
%
range decreases with increasing $Q^2$. This is to be expected since the
resolved contribution diminishes with increasing $Q^2$, so that the NLO
corrections to the resolved cross section and the effect of the evolution of
the photon PDF diminish as well. Similar conclusions can be drawn from Fig.\
\ref{fig:6}, where we plotted $\d\sigma/\d z_{\p}$ in the four
%
%
$Q^2+p_T^{*2}$-ranges. For all four ranges the dependence on $M_{\gamma}$ is
small if $z_{\p} \geq 0.4$. Only for the two lowest $Q^2+p_T^{*2}$-ranges
this cross section depends on $M_{\gamma}$ for the two lowest $z_{\p}$-bins
and is strongest for $z_{\p}\in [0;0.2]$ and $Q^2+p_T^{*2}\in [20;35]$
GeV$^2$. \\

As is well known the photon PDFs consist of a point-like or anomalous part,
the hadron-like part and the gluon part. The compensation of the
$M_{\gamma}$-dependence between the NLO direct and the LO resolved cross
section occurs via the anomalous part. This means that this part of the PDFs
is closely related to the direct cross section. If one assumes that the
direct part obeys factorization , {\it i.e.} has no suppression factor, it
was suggested in \cite{Klasen:2004qr}, that one possibility to avoid the
non-compensation of the $M_{\gamma}$-dependence between the suppressed
resolved cross section and the unsuppressed direct cross section would be to
suppress only the hadron-like and the gluon part of the PDFs. Fortunately,
in the SaS-type photon PDFs the hadron-like and the anomalous part and the
effect of their evolution are presented separately, so that the effect of
the suppression of the hadron-like part can be investigated for all scales.
It turns out, however, that the hadron-like part contributes only a very
small fraction to the total resolved cross section. For $Q^2=5$ GeV$^2$,
this fraction is $3\cdot 10^{-3}$, and it decreases strongly with increasing
$Q^2$. This means that the anomalous component dominates the total resolved
cross section and a suppression of the hadron-like part alone would not be
sufficient to account for the experimental data at low $Q^2$. Although in
the case of photoproduction ($Q^2 \simeq 0$) the hadron-like part is
supposed to be larger, it would be quite artificial to have two different
mechanisms for the $M_{\gamma}$-scale compensation depending on whether
$Q^2 > 0$ or $Q^2 \simeq 0$. For this reason we prefer the mechanism with
the suppression in the DIR$_{\rm IS}$ contribution to the NLO corrections of
the direct cross section, which works for all $Q^2$.

\section{Conclusions}
\label{sec:3}

In summary, we propose in this paper a new factorization scheme for
diffractive production of jets in low-$Q^2$ deep inelastic scattering.
By suppressing not only the resolved photon contribution, but also the
unresummed logarithm as well as scheme-dependent finite terms in the NLO
direct initial state correction, factorization scheme and scale invariance
is restored up to higher order effects, while at the same time the cut-off
invariance required in phase space slicing methods is preserved. \\

For pedagogical reasons, we have chosen in this paper the kinematical region
of finite, but low photon virtuality $Q^2$, which exposes and regularizes a
logarithmic virtual photon initial state singularity. We do, however, not
rely on the finiteness of $Q^2$, but rather separate suppressed and
unsuppressed terms using the hard transverse momentum scale $p_T^*$, so that
our scheme is equally valid for real photoproduction. Furthermore, due to
the universality of the initial state singularity, our new factorization
scheme should apply to any other diffractive photon-induced process with a
second hard scale, such as inclusive hadron-production at large $p_T^*$ or
heavy quark and quarkonium production. \\

The scheme- and scale invariance has been demonstrated numerically using
the kinematics of a recent H1 analysis, differential in $Q^2$ and parton
momentum fraction in the pomeron $z_{\p}$. Very good stability with respect
to scheme- and scale variations and good agreement with the experimental
data has been found. \\

Individual studies of the anomalous and hadronic component of the resolved
virtual photon show that the parton densities in the virtual photon are
dominated by the resummed pointlike component, as expected from
non-diffrac\-tive deep inelastic scattering. While suppressing only the
hadronic component in the photon PDFs would also be scheme- and
scale-invariant and therefore be theoretically consistent, our numerical
study shows that this alternative is thus phenomenologically not viable.




\end{document}